%% file: tc_reservoir_letter.tex
\IfFileExists{revtex4-2.cls}{%
  \documentclass[prl,twocolumn,superscriptaddress,showpacs]{revtex4-2}\def\REVTEX{}}{%
  \documentclass[twocolumn,10pt]{article}\usepackage[margin=0.75in]{geometry}}
\ifdefined\REVTEX\else
  \providecommand{\affiliation}[1]{\\{\small\itshape #1}}\providecommand{\email}[1]{}
  \newenvironment{acknowledgments}{\paragraph*{Acknowledgments.}}{}
\fi
\usepackage{graphicx,amsmath,amssymb,xcolor,booktabs,float,adjustbox}
\providecommand{\onecolumngrid}{\onecolumn}
\usepackage[colorlinks=true,allcolors=blue]{hyperref}

\ifdefined\REVTEX\else\renewcommand{\affiliation}[1]{\date{\small\itshape #1}}\fi
\begin{document}
\title{Where hurricanes get their spin: angular momentum versus kinetic energy}
\ifdefined\REVTEX
\author{Boris Galperin}
\email{bgalperin@usf.edu}
\affiliation{College of Marine Science, University of South Florida, St.~Petersburg, Florida, USA}
\author{Jun A. Zhang}
\email{jun.zhang@miami.edu}
\affiliation{Department of Atmospheric Sciences, University of Miami, Miami, Florida, USA}
\else
\author{Boris Galperin$^{1}$ and Jun A. Zhang$^{2}$}
\date{\small\itshape $^1$College of Marine Science, University of South Florida, St.~Petersburg, Florida, USA\\$^2$Department of Atmospheric Sciences, University of Miami, Miami, Florida, USA}
\fi
\ifdefined\REVTEX\else\maketitle\fi
\begin{abstract}
Tropical cyclones almost never form within a few degrees of the equator. Their energy source, the heat 
engine fueled by the warm ocean, is available there as everywhere else. What is missing must be the other ingredient a cyclone needs, absolute angular momentum about the local vertical. Across reconnaissance intensity categories, characteristic wind energy increases and the radius of maximum wind decreases, while the characteristic specific absolute angular momentum $M$ varies comparatively little; energy and absolute angular momentum are thus supplied separately. We propose that $M$ is drawn from the mesoscale turbulence in which storms form, and measure the rotational content of this turbulence in ERA5 winds over the western Pacific. Including the storms, the rotational part of the one-dimensional spectrum along latitude circles follows the Coriolis branch $Cf^2k^{-3}$ of rotating stratified turbulence, with $C=0.047$, about a third of the theoretical value, from $12^\circ$ to $24^\circ$N. Between storms, this value is lower and grows, approximately, as $f$;
we interpret the storms as its magnified, predominantly cyclonic tail. Toward the equator the energy of the turbulence persists but its rotational content collapses: equatorward of $\sim6^\circ$N, more than one third
of the energy is divergent. Typhoon Vamei, which formed at $1.5^\circ$N in December 2001, was 
preceded by rotational variance above the 99th percentile of December climatology.
\end{abstract}
\ifdefined\REVTEX\maketitle\fi

\textit{Introduction.}---Tropical cyclogenesis requires low-level absolute vorticity \cite{Gray1968}, and cyclones almost never form within $5^\circ$ of the equator. Typhoon Vamei (2001), at $1.5^\circ$N, is the closest exception on record \cite{Changetal2003}. The classical argument is that ambient absolute vorticity must 
supply the storm's absolute angular momentum \cite{Emanuel2003}, and that as $f \to 0$, horizontal convergence can no longer generate relative vorticity \cite{Changetal2003}. Nor do cyclones cross the
equator. This is often attributed to the change of sign of $f$, under which inflow conserving absolute 
angular momentum would spin the vortex down, and to the poleward $\beta$-drift. We argue that the 
obstacle is not the sign of $f$ but its zero: the rotational content of the environment grows with $|f|$, is the same in both hemispheres, and collapses to a floor several degrees before the equator. A storm never
arrives to find out that $f$ has changed sign. The vorticity of a mature storm is assembled upscale, as convective vortical hot towers merge \cite{Montgomeryetal2006} inside easterly-wave ``marsupial pouches''
\cite{Dunkertonetal2009}. In this paradigm, rotating deep convection grows in a cyclonic, rotation-rich environment \cite{Montgomeryetal2006,MontgomerySmith2017}. What makes an environment rotation-rich,
how rich it must be, and why that richness depends on latitude have not been quantified. The rotation-rich pouches of this paradigm are fluctuations of a rotational background; we measure the statistics of that background, of which, we suggest, the storms are the magnified tail. We propose that this store is the mesoscale turbulence of the atmosphere, present it as an environmental 
diagnostic that can be measured, and show that precisely its rotational content, not its energy, collapses toward the equator. 
How the storm draws on it remains a hypothesis.

\textit{Energy versus absolute angular momentum.}---Energy and absolute angular momentum are conserved independently, because they are dictated by different symmetries \cite{LandauLifshitz1976}; a tropical cyclone needs a supply of each. The spin of a vortex about the local vertical is measured by its absolute angular momentum $M=rv+\tfrac12 f r^2$, as sustained by the absolute vorticity $f+\zeta$. Only the vertical component of the Earth's rotation, $f = 2\Omega\sin\phi$, contributes. At the equator, 
the rotation vector lies in the horizontal 
plane: $f \to 0$, while $\tilde f = 2\Omega\cos\phi$ reaches its maximum, $2\Omega$. Rotation organizes turbulence into columns along its axis \cite{Cambonetal1997}; at the equator these columns would lie horizontally, which the stable stratification suppresses. Turbulence there can keep its energy, in horizontal-axis motions and equatorial waves, while losing
the vertical vorticity a cyclone needs. Energy and absolute angular momentum must therefore be diagnosed separately, by splitting the kinetic-energy spectrum into its rotational (vertical-vorticity) and divergent parts.

\textit{The reservoir.}---Above the mesoscale, atmospheric spectra follow a $k^{-3}$ power law \cite{NastromGage1985}. The classical reading is Charney's enstrophy cascade, $E=C\eta^{2/3}k^{-3}$, whose level is set by the enstrophy flux, $\eta$, 
blind to latitude \cite{Charney1971}. In a laboratory annulus at fixed rotation, the level scales with $N^2$ \cite{Dingetal2026}.
Structure-function analysis of aircraft data, however, indicated that the Coriolis term, not nonlinear advection, may control the 
$k^{-3}$ range at 100--1000~km, and its level may strongly depend on latitude \cite{LindborgCho2001}. If this is the case, then, with $E_h(k_h)$ the horizontal kinetic-energy spectrum normalized such that $\int E_h\,dk_h = \tfrac12\langle u^2+v^2\rangle$ and $k_h$ the horizontal wavenumber, the product $E_h(k_h)k_h^3$, whose units are s$^{-2}$, can depend only on $f$ in this subrange. This precludes a dependence on the horizontal energy flux, $\Pi_\varepsilon$ \cite{Galperinetal2025}, and points to the scaling $E_h(k_h) \propto f^2 k_h^{-3}$. Such dependence in rotating turbulence was derived analytically using quasi-normal 
scale elimination (QNSE) \cite{SukorianskyGalperin2016,Galperinetal2021},     
\begin{equation}
E_h(k_h)=C_h f^2 k_h^{-3},     \label{eq:Ef}
\end{equation}
with the nondimensional coefficient $C_h\approx0.33$. Equation \eqref{eq:Ef} reproduces the Nastrom--Gage 
spectrum \cite{GalperinSukoriansky2020} and the corresponding vorticity variance is
\begin{equation}
\langle\zeta^2\rangle=2\int k_h^2E_h\,dk_h=2C_h f^2\ln\frac{k_{\max}}{k_{\min}}. \label{eq:z2}
\end{equation}
One infers that $\zeta_{\rm rms}$ is of order $f$, independent of $\Pi_\varepsilon$: planetary absolute angular 
momentum is held in turbulent form, at eddy Rossby number of order one. In Charney's form, a dependence on $f$ can
enter only through the forcing, via $\eta$; Eq.~(\ref{eq:Ef}), on the other hand, requires no assumption about the forcing,
which in the tropics is convective rather than baroclinic. The data can decide between the two readings.

\textit{Hurricanes gain energy at nearly constant absolute angular momentum.}---Inside a tropical cyclone, rotation is dominated
by the vortex itself. Writing $M(r)=U(r)r+\tfrac12 f r^2=\tfrac12 r^2\tilde f(r)$ defines an effective Coriolis parameter, $\tilde f=2M/r^2$. 
Reconnaissance spectra of storms of all categories follow Eq.~(\ref{eq:Ef}) with $f$ replaced by $\tilde f$, with the same 
constants as in the free troposphere. At the edge of the vortex, $\tilde f\to f$, and they merge with the Nastrom--Gage spectra
\cite{Galperinetal2025}. The category means of that data set show what intensification does to $M\simeq U_{\max}R_{mw}$: from tropical storm to category 5 it stays between $1.6$ and $2.3\times10^{6}$~m$^2$~s$^{-1}$, while $\Pi_\varepsilon$ in the inner region ($r\le2R_{mw}$), where most of the damage is done, grows
fifteenfold, $U_{\max}^2$ grows elevenfold, and $R_{mw}$ shrinks from 82 to 25~km \cite{Galperinetal2025}. The category
means are consistent with intensification by contraction at approximately constant characteristic $M$, as the turbulence evolves
from peristrophic to zonostrophic and the flow annularizes, with $R_{mw}$ marking the end of the zonostrophic range \cite{Galperinetal2025}. Vortex energy is gathered from the environment's large scales into a core of a few tens of kilometres, and losses grow far faster than the winds: normalized U.S. damage per storm rises by a factor of about 250 from category 1 to category 4 \cite{PielkeLandsea1998}.
In the category composites the peristrophic coefficients fall with intensity, the longitudinal one from 0.040 in tropical depressions
to 0.007 at category 5, while the zonostrophic coefficient rises from 0 to 4.5 \cite{Galperinetal2025}, consistent with peristrophic turbulence being converted into the annular jet of the eyewall. The starting value is consistent with the environment: in the same convention, that of the one-dimensional longitudinal spectrum normalized to $\langle u^2\rangle$, for which QNSE gives 0.0926 \cite{Galperinetal2021}, the ERA5 spectra below give a longitudinal coefficient of 0.031 at $12^\circ$--$24^\circ$N including the storms and 0.022 between them. The heat engine \cite{Emanuel1986} supplies the energy, and its maximum intensity does not depend on $f$; but energy alone could be deposited at any scale, including in an expanding vortex. It is the conservation of $M$ that channels it: with $M$ fixed, the kinetic energy at the radius of maximum wind is $\sim M^2/R_{mw}^2$, so added energy is most readily absorbed by contraction, with $U_{\max}\simeq M/R_{mw}$. The engine sets how much energy; absolute angular momentum sets where it goes. Individual storms differ in size and hence in $M$, but $M$ itself must be drawn from the environment. Far from the center $M\simeq\tfrac12(f+\zeta_{\rm env})r_0^2$, so a typical $M=2\times10^{6}$~m$^2$~s$^{-1}$ supplied by $f$ alone requires a catchment radius $r_0=(2M/f)^{1/2}$ of 280~km at $20^\circ$N, 560~km at $5^\circ$N and 1020~km at $1.5^\circ$N. Toward the equator only the relative vorticity of the environment can supply it, which is what the rotational reservoir diagnoses. Reconnaissance flight segments are too short to resolve the $k^{-3}$ range, so we measure the reservoir in reanalysis.

\textit{Data and method.}---We use ERA5 \cite{Hersbachetal2020} winds at 850 and 700~hPa over the western Pacific
($5^\circ$S--$45^\circ$N, $100^\circ$--$180^\circ$E, $0.25^\circ$), daily at 00~UTC for August--October 2019--2021 (276 days), and over the central Pacific ($30^\circ$S--$30^\circ$N, $160^\circ$E--$140^\circ$W) for August--October 2019--2021 and January--March 2020--2022. Along each latitude circle the zonal wind $u$ is the longitudinal and the meridional wind $v$ the transverse component. Spectra are computed after removing a linear trend and applying a Hann window, and averaged in $2^\circ$ bands. The one-dimensional Helmholtz decomposition of B\"uhler \emph{et al.} \cite{Buhleretal2014} splits $E_L+E_T$ into additive rotational and divergent parts, $E_\psi+E_\phi=E_L+E_T$. This split is the horizontal projection of the Craya--Herring decomposition of a three-dimensional incompressible field into toroidal (vortical) and poloidal (wave) components \cite{Craya1958,Herring1974}. Here $E_L$ and $E_T$ are the one-dimensional spectra of $u$ and $v$ along latitude circles in the zonal wavenumber $k$, normalized so that $\int E_L\,dk=\tfrac12\langle u^2\rangle$ and $\int E_T\,dk=\tfrac12\langle v^2\rangle$. For a horizontally isotropic, non-divergent field obeying Eq.~(\ref{eq:Ef}), $E_\psi=\tfrac12C_hf^2k^{-3}$; we quote the measured level as $C=E_\psi k^3/f^2$, so that $C_h=2C$, and the QNSE value corresponds to $C\approx0.17$. For a $k^{-n}$ range, a horizontally non-divergent field gives $E_T/E_L=n$ and a purely divergent one $1/n$. These limits are kinematic: the split concerns the horizontal field on a level and requires only horizontal isotropy, not two-dimensionality. A three-dimensional solenoidal field has horizontal divergence $-\partial w/\partial z$; isotropic turbulence gives $E_T/E_L=(n+1)/2$ \cite{MoninYaglom1975}, and rotating turbulence, in the QNSE theory, gives $E_T/E_L\approx2.6$ for $n=3$ \cite{GalperinSukoriansky2020}. Levels are the median of $E k^3$ over wavelengths 600--2500~km, well within the scales ERA5 resolves. Horizontal isotropy holds poleward of $14^\circ$N (zonal and meridional transects agree within 25\%) but not near the equator, where meridional levels are 1.5--2 times the zonal ones. We therefore also split the energy with two-dimensional band-passed (600--2000~km) vorticity and divergence, which requires no isotropy.

\begin{figure*}[t]
\centering
\includegraphics[width=\textwidth]{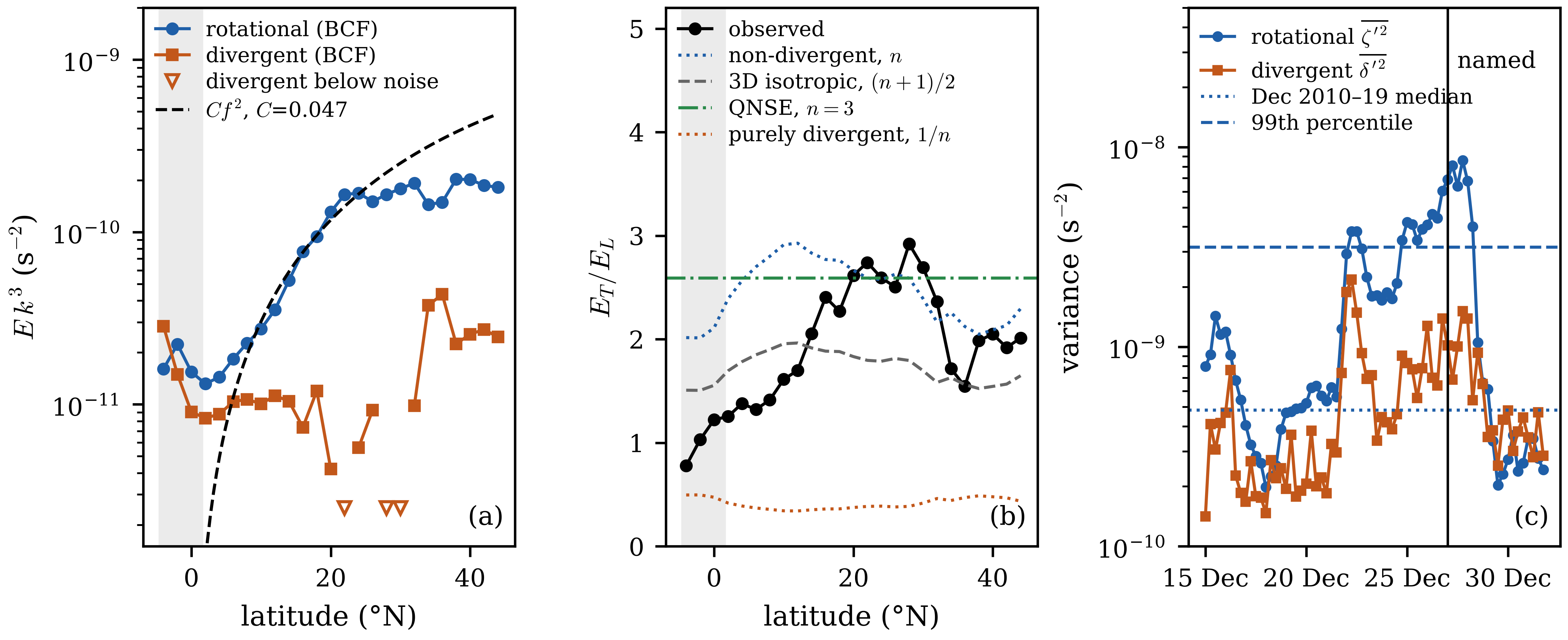}
\caption{(a) Rotational and divergent one-dimensional levels $E_\psi k^3$ and $E_\phi k^3$ at 600--2500~km, ERA5 850~hPa, western Pacific, August--October 2019--2021; dashed: $C f^2$ fitted over $12^\circ$--$24^\circ$N. Shading: latitudes crossing the Maritime Continent. (b) Ratio of transverse to longitudinal spectra, with the two-dimensional non-divergent ($n$) and purely divergent ($1/n$) limits and the isotropic three-dimensional solenoidal value $(n+1)/2$ for the local slope $k^{-n}$; dash-dotted: QNSE value for rotating turbulence with $n=3$ \cite{GalperinSukoriansky2020}. (c) Mesoscale rotational and divergent variance at 850~hPa in the box $0^\circ$--$5^\circ$N, $102^\circ$--$112^\circ$E during December 2001, with the median and 99th percentile of December 2010--2019; vertical line: Vamei named.}
\label{fig:era5}
\end{figure*}

\textit{The rotational reservoir follows $f^2$.}---From $12^\circ$ to $24^\circ$N, the latitudes where most western Pacific storms form, the rotational level follows $E_\psi=Cf^2k^{-3}$, the one-dimensional form of Eq.~(\ref{eq:Ef}), with $C=0.047$ (rms scatter 12\%) while $f^2$ changes by a factor of 3.8 [Fig.~\ref{fig:era5}(a)], and the spectral slope is $-2.8$ to $-2.9$. This corresponds to $C_h=2C\approx0.09$, about a third of the QNSE value. ERA5 at 250~hPa over the midlatitude Pacific, the region of the Nastrom--Gage observations, gives $C_h\approx0.1$--$0.2$, a factor of 2--3 below the aircraft spectra that QNSE reproduces \cite{GalperinSukoriansky2020}; the reanalysis thus underestimates the level, by a comparable factor at 250~hPa in the tropics and in midlatitudes; a bias that is similar at all latitudes would not affect the dependence on latitude \cite{SM}. The two-dimensional band-passed vorticity, which requires no isotropy, gives the same result over $12^\circ$--$24^\circ$N (scatter 13\%); at $8^\circ$--$10^\circ$N, where isotropy is weaker, it lies 30--60\% above $Cf^2$. The law holds for the basin as a whole; east of $130^\circ$E alone the level falls by about a factor of two between $12^\circ$ and $24^\circ$N. Of the band-passed vorticity variance at $12^\circ$--$24^\circ$N, 68--77\% is cyclonic: the reservoir is predominantly cyclonic, although variance alone does not show transport into the storm. This level includes the tropical cyclones present during the period. Masking a disc of 500~km radius around each storm detected in the reanalysis (0.8 per day, 4\% of the area) leaves the level nearly unchanged at $12^\circ$N but halves it at $20^\circ$--$24^\circ$N, so that between storms it grows only about as $f$; masking discs of the same number, radius and latitudes at random longitudes changes nothing. We interpret the storms not as an addition to the background but as its fluctuations, magnified where the rotation is strong, so that the $f^2$ law describes the background together with this magnified tail; the masking establishes their contribution to the scaling, not how they develop from the background. Equatorward of $10^\circ$N storms are nearly absent, and the results below are unaffected by the masking. Tests of the scale range, sampling uncertainty and storm masking are given in the Supplemental Material \cite{SM}. Over the same band $E_T/E_L$ rises from 1.4 at $8^\circ$N to 2.4--2.7 at $16^\circ$--$24^\circ$N, close to the QNSE value for rotating turbulence and well above the isotropic three-dimensional value [Fig.~\ref{fig:era5}(b)]; between $20^\circ$ and $30^\circ$N the divergent part falls below what the decomposition can resolve. The enstrophy-cascade reading offers no reason for this latitude dependence. Oceanic mesoscale spectra show the same dependence of the spectral level on $f$ \cite{GalperinSukoriansky2020,Galperinetal2021}. Poleward of $26^\circ$N the rotational level falls below $Cf^2$. The peristrophic coefficients decrease with latitude much as they decrease with storm category: in the convention of Ref.~\cite{Galperinetal2025}, the longitudinal coefficient falls from 0.031 at $16^\circ$N to 0.013 at $44^\circ$N, while in the storm-category composites it falls from 0.040 in tropical depressions to 0.007 at category 5 \cite{Galperinetal2025}. Stronger rotation, planetary or cyclostrophic, depletes the peristrophic range as the flow becomes zonostrophic, so Eq.~(\ref{eq:Ef}) with constant $C$ holds only over a limited range of rotation, here the tropics and subtropics. Consistently, the law appears where storms form: over the central Pacific, the rotational level again follows $Cf^2$ in the South Pacific cyclone season (January--March), with $C\approx0.033$ from $10^\circ$ to $20^\circ$S, but not in the quiet North Pacific or in the austral winter, where the spectra are shallower than $k^{-3}$ and the level falls below $Cf^2$.

\textit{Energy with reduced rotational content.}---Equatorward of $\sim6^\circ$N the rotational level stops following $f^2$ and settles on a floor of $\approx1.5\times10^{-11}$~s$^{-2}$. The divergent level, $\approx1\times10^{-11}$~s$^{-2}$, is nearly independent of latitude from the equator to $18^\circ$N. The kinetic energy therefore levels off rather than vanishing, but its composition changes: $E_T/E_L$ drops to 1.2--1.4, below even the isotropic three-dimensional value $(n+1)/2\approx1.6$--$1.9$ for the local slopes (the observed ratio falls below that value equatorward of $13^\circ$N), and the divergent part grows from a negligible fraction to more than a third of the total, about twice what isotropic solenoidal turbulence carries. The excess is therefore unlikely to be explained by the $\partial w/\partial z$ of three-dimensional turbulence alone. The two-dimensional fields, free of the isotropy assumption, give the same result: over the central Pacific the divergent share is 0.36--0.42 within $6^\circ$ of the equator, while over the western Pacific it falls to 0.05--0.07 at $18^\circ$--$26^\circ$N, close to the $\approx0.06$ implied by the QNSE ratio for $n=3$. The energy persists; its rotational content about the vertical does not. This is the tilt of the rotation vector made visible in the spectra. This is consistent with the Coriolis branch itself: its crossover to the flux-controlled range occurs at 
$L_c \sim (\Pi_\varepsilon/f^3)^{1/2}$ \cite{Galperinetal2021}, which grows without bound as $f\to0$. Near the equator the 
large scales therefore pass from the Coriolis branch, whose level is set by $f$, to a range whose level is set by the energy flux and which has reduced rotational content. Near the equator the total spectrum is therefore controlled by the energy flux; the rotational part falls to a finite floor. Over the central Pacific the floor is symmetric about the equator in both seasons, with $E_T/E_L\approx1$ at the equator and the same level, $\approx1\times10^{-11}$~s$^{-2}$, as in the western Pacific. Satellite altimetry shows the same transition in the ocean: sea-surface-height spectra steeper than $k^{-2}$ everywhere poleward of $\sim20^\circ$ give way to flatter spectra at low latitudes, where the motions are no longer geostrophically balanced \cite{XuFu2012}. Aquaplanet simulations place the minimum genesis distance from the equator at the equatorial Rhines scale \cite{ChavasReed2019}; with $U\approx10$~m~s$^{-1}$ this is $\approx650$~km, close to where the rotational level leaves $Cf^2$. The spectra show what fills that equatorial belt: divergent energy over a depleted rotational reservoir. Such planetary scales describe the environment; inside the vortex, rotation is cyclostrophic, and the relevant parameters are the cyclostrophic counterparts of $f$ and $\beta$ \cite{Galperinetal2025}.

\textit{Vamei.}---In the box $0^\circ$--$5^\circ$N, $102^\circ$--$112^\circ$E we compute the variance of vorticity and divergence after removing scales larger than $\sim450$~km, and compare it with every day of December 2010--2019 [Fig.~\ref{fig:era5}(c)]. The rotational variance exceeded the 99th percentile of the climatology on 22 December 2001, as the Borneo vortex moved over the narrow equatorial sea during an intense and persistent cold surge \cite{Changetal2003}, dropped to about four times its median (3.6--4.7) on 23--24 December, and rose again from 25 December as the storm developed in the box; over 22--24 December the rotational-to-divergent ratio rose from 1.5 to 4.5. After the storm left the box the variance fell straight back to the floor. A box average cannot separate the vortex from its environment, and Vamei is a single case, but it is consistent with the proposed interpretation: genesis at the equator when the rotational content is lifted far above its floor, here by the surge, which Chang \emph{et al.} identify as the source of the background cyclonic vorticity \cite{Changetal2003}. In Earth-like simulations, too, genesis comes closest to the equator over the Maritime Continent \cite{ChavasReed2019}.

\textit{Implications.}---Empirical predictors of cyclogenesis, fitted to the observed seasonal cycle and geography of storm formation, take their rotational ingredient from the monthly-mean absolute vorticity, which near the equator reduces to $f$ \cite{EmanuelNolan2004}. The rotational reservoir follows $f^2$ in the mean, but it is a mesoscale quantity that responds within days to surges, waves and seasons, and it can be measured. Near the equator the relevant question becomes how often, and by how much, the rotational part rises above the divergent floor. That question bears directly on the coasts of Southeast Asia, where Vamei made landfall just north of Singapore.

\textit{Data availability.}---The ERA5 data are publicly available from the Copernicus Climate Change Service \cite{Hersbachetal2020}. The reconnaissance composites are those of Ref.~\cite{Galperinetal2025}. The analysis scripts and the derived spectra will be deposited in the USF Digital Commons data repository upon acceptance and are available from the authors on request until then.

\begin{acknowledgments}
ERA5 data were obtained from the Copernicus Climate Change Service. Data-analysis code, figures, and portions of the text were developed with assistance from Claude Opus 5.5 (Anthropic). The authors directed the analysis, verified the results, and take full responsibility for the content.
\end{acknowledgments}
\bibliographystyle{apsrev4-2}
\bibliography{HurricaneBib}
% Supplemental Material appended for the arXiv version
\clearpage
\onecolumngrid
\setcounter{table}{0}\setcounter{figure}{0}\setcounter{equation}{0}
\renewcommand{\thetable}{S\arabic{table}}\renewcommand{\thefigure}{S\arabic{figure}}\renewcommand{\theequation}{S\arabic{equation}}
\input{tc_sm_body}
\end{document}

%% file: tc_sm_body.tex
\begin{center}{\large\bfseries Supplemental Material: robustness of the rotational level}\\[4pt]
\textit{Where hurricanes get their spin: angular momentum versus kinetic energy}\end{center}

All tests use the rotational level of the one-dimensional Helmholtz decomposition, ERA5 850~hPa, western Pacific ($100^\circ$--$180^\circ$E), August--October 2019--2021 (276 days), averaged over $12^\circ$--$24^\circ$N in $2^\circ$ bands. $C$ is the mean of $E_\psi k^3/f^2$, where $E_\psi$ is the rotational part of the one-dimensional spectrum along latitude circles, normalized so that $\int(E_L+E_T)\,dk=\tfrac12\langle u^2+v^2\rangle$ (the corresponding horizontal-wavenumber coefficient is $C_h=2C$; QNSE gives $C_h\approx0.33$); the exponent $p$ is the slope of $\log E_\psi k^3$ against $\log f^2$ ($p=1$ is the $f^2$ law).

\section*{S1. Isotropy-free analysis}
Two-dimensional band-passed (600--2000~km) vorticity and divergence in boxes $12^\circ$ tall, energy-weighted by $K^{-2}$. Over $12^\circ$--$24^\circ$N the rotational level divided by $f^2$ has an rms scatter of 13\% (100--180$^\circ$E). At $8^\circ$--$10^\circ$N it lies 30--60\% above $Cf^2$. East of $130^\circ$E alone the level falls by a factor of $\approx2$ between $12^\circ$ and $24^\circ$N. The cyclonic share of band-passed vorticity variance is 0.68--0.77 at $12^\circ$--$24^\circ$N.

\section*{S2. Scale range}
\begin{center}\begin{tabular}{lccc}\toprule
Wavelength band (km) & $C$ & rms scatter & $p$\\\midrule
600--2500 (main text) & 0.047 & 0.11 & 1.22\\
800--2500 & 0.046 & 0.08 & 1.10\\
1000--2500 & 0.046 & 0.08 & 1.06\\
600--1500 & 0.048 & 0.13 & 1.25\\
500--2000 & 0.048 & 0.11 & 1.22\\
600--3000 & 0.047 & 0.11 & 1.21\\\bottomrule\end{tabular}\end{center}

\section*{S3. Sampling uncertainty}
Block bootstrap over the nine months (1000 draws): $C=0.047$ (5--95\%: 0.039--0.055), $p=1.21$ (0.87--1.52).

\section*{S4. Tropical cyclones}
Storms are detected in ERA5 as local maxima of 850-hPa vorticity smoothed over $\approx1^\circ$, exceeding the threshold, with maximum wind within $2^\circ$ above the threshold. A disc of the given radius around each centre is removed with a $2^\circ$ cosine taper; the rest of each latitude circle is kept, and spectra are normalized by the mean squared window. In the controls, each storm disc is moved to a random longitude at least $10^\circ$ from every detected storm on the same day, keeping its latitude and radius, so that the removed area and its latitude distribution match the storm masks.
\begin{center}\small\begin{tabular}{clcccc}\toprule
& Mask & centres/day & area & $C$ & $p$\\\midrule
1 & none & -- & 0 & 0.047 & 1.22\\
2 & $\zeta>3\times10^{-4}$, $U>20$, $r=5^\circ$ & 0.80 & 4\% & 0.028 & 0.55\\
3 & $\zeta>1.5\times10^{-4}$, $U>15$, $r=5^\circ$ & 2.47 & 9\% & 0.024 & 0.53\\
4 & $\zeta>1.5\times10^{-4}$, $U>17$, $r=8^\circ$ & 2.13 & 16\% & 0.021 & 0.37\\
5 & control for row 2 & 0.80 & 4\% & 0.049 & 1.20\\
6 & control for row 3 & 2.47 & 8\% & 0.052 & 1.19\\\bottomrule\end{tabular}\end{center}
\noindent $\zeta$ in s$^{-1}$, $U$ in m~s$^{-1}$. Controls: same number of discs, radius and latitudes as the storm mask, at random longitudes at least $10^\circ$ from any detected storm; the removed areas of storm masks and controls differ by about one percentage point (rows 3 and 6), well below what would matter for $C$.

With the strict mask, $C$ falls from 0.036 at $12^\circ$N to 0.021 at $24^\circ$N; the full-field values are 0.039 and 0.048. Equatorward of $10^\circ$N the rotational and divergent levels, the divergent share (0.36--0.39 at $0^\circ$--$6^\circ$N) and $E_T/E_L$ (1.2--1.4) are unchanged by masking.
\section*{S5. Absolute level of the reanalysis}
ERA5 $u$, $v$ at 250 and 850~hPa over the midlatitude Pacific ($25^\circ$--$55^\circ$N, $140^\circ$E--$140^\circ$W) and at 250~hPa over the western Pacific box of the main text; daily 00~UTC; same spectral method and normalization. $C_h=2C$; QNSE gives $C_h\approx0.33$, in agreement with aircraft (Nastrom--Gage) spectra.
\begin{center}\begin{tabular}{lcc}\toprule
Region, level, season & 600--2500 km & 1200--3000 km\\\midrule
Midlatitude Pacific, 250 hPa, Aug--Oct & 0.06--0.18 & 0.09--0.20\\
Midlatitude Pacific, 250 hPa, Jan--Mar & 0.045--0.07 & 0.08--0.11\\
Midlatitude Pacific, 850 hPa, all seasons & 0.01--0.08 & --\\
Western Pacific $12^\circ$--$24^\circ$N, 250 hPa, Aug--Oct & 0.14--0.18 & 0.12--0.16\\
Western Pacific $12^\circ$--$24^\circ$N, 850 hPa, Aug--Oct (main text) & 0.09 & --\\\bottomrule\end{tabular}\end{center}
Even at the altitude and in the region of the aircraft observations ERA5 lies a factor of 2--3 below them; winter spectra steepen to $k^{-3.5}$--$k^{-3.9}$ at 600--1200~km, consistent with damping near the effective resolution. The factor is common to all latitudes and does not affect the $f^2$ dependence.

%% file: tc_reservoir_letter.bbl
%apsrev4-2.bst 2019-01-14 (MD) hand-edited version of apsrev4-1.bst
%Control: key (0)
%Control: author (72) initials jnrlst
%Control: editor formatted (1) identically to author
%Control: production of article title (-1) disabled
%Control: page (0) single
%Control: year (1) truncated
%Control: production of eprint (0) enabled
\begin{thebibliography}{27}%
\makeatletter
\providecommand \@ifxundefined [1]{%
 \@ifx{#1\undefined}
}%
\providecommand \@ifnum [1]{%
 \ifnum #1\expandafter \@firstoftwo
 \else \expandafter \@secondoftwo
 \fi
}%
\providecommand \@ifx [1]{%
 \ifx #1\expandafter \@firstoftwo
 \else \expandafter \@secondoftwo
 \fi
}%
\providecommand \natexlab [1]{#1}%
\providecommand \enquote  [1]{``#1''}%
\providecommand \bibnamefont  [1]{#1}%
\providecommand \bibfnamefont [1]{#1}%
\providecommand \citenamefont [1]{#1}%
\providecommand \href@noop [0]{\@secondoftwo}%
\providecommand \href [0]{\begingroup \@sanitize@url \@href}%
\providecommand \@href[1]{\@@startlink{#1}\@@href}%
\providecommand \@@href[1]{\endgroup#1\@@endlink}%
\providecommand \@sanitize@url [0]{\catcode `\\12\catcode `\$12\catcode
  `\&12\catcode `\#12\catcode `\^12\catcode `\_12\catcode `\%12\relax}%
\providecommand \@@startlink[1]{}%
\providecommand \@@endlink[0]{}%
\providecommand \url  [0]{\begingroup\@sanitize@url \@url }%
\providecommand \@url [1]{\endgroup\@href {#1}{\urlprefix }}%
\providecommand \urlprefix  [0]{URL }%
\providecommand \Eprint [0]{\href }%
\providecommand \doibase [0]{https://doi.org/}%
\providecommand \selectlanguage [0]{\@gobble}%
\providecommand \bibinfo  [0]{\@secondoftwo}%
\providecommand \bibfield  [0]{\@secondoftwo}%
\providecommand \translation [1]{[#1]}%
\providecommand \BibitemOpen [0]{}%
\providecommand \bibitemStop [0]{}%
\providecommand \bibitemNoStop [0]{.\EOS\space}%
\providecommand \EOS [0]{\spacefactor3000\relax}%
\providecommand \BibitemShut  [1]{\csname bibitem#1\endcsname}%
\let\auto@bib@innerbib\@empty
%</preamble>
\bibitem [{\citenamefont {Gray}(1968)}]{Gray1968}%
  \BibitemOpen
  \bibfield  {author} {\bibinfo {author} {\bibfnamefont {W.~M.}\ \bibnamefont
  {Gray}},\ }\href
  {https://doi.org/10.1175/1520-0493(1968)096%3C0669:GVOTOO%3E2.0.CO;2}
  {\bibfield  {journal} {\bibinfo  {journal} {Mon.\ Wea.\ Rev.}\ }\textbf
  {\bibinfo {volume} {96}},\ \bibinfo {pages} {669} (\bibinfo {year}
  {1968})}\BibitemShut {NoStop}%
\bibitem [{\citenamefont {Chang}\ \emph {et~al.}(2003)\citenamefont {Chang},
  \citenamefont {Liu},\ and\ \citenamefont {Kuo}}]{Changetal2003}%
  \BibitemOpen
  \bibfield  {author} {\bibinfo {author} {\bibfnamefont {C.-P.}\ \bibnamefont
  {Chang}}, \bibinfo {author} {\bibfnamefont {C.-H.}\ \bibnamefont {Liu}},\
  and\ \bibinfo {author} {\bibfnamefont {H.-C.}\ \bibnamefont {Kuo}},\ }\href
  {https://doi.org/10.1029/2002GL016365} {\bibfield  {journal} {\bibinfo
  {journal} {Geophys. Res. Lett.}\ }\textbf {\bibinfo {volume} {30}},\ \bibinfo
  {eid} {1150} (\bibinfo {year} {2003})}\BibitemShut {NoStop}%
\bibitem [{\citenamefont {Emanuel}(2003)}]{Emanuel2003}%
  \BibitemOpen
  \bibfield  {author} {\bibinfo {author} {\bibfnamefont {K.~A.}\ \bibnamefont
  {Emanuel}},\ }\href {https://doi.org/10.1146/annurev.earth.31.100901.141259}
  {\bibfield  {journal} {\bibinfo  {journal} {Annu. Rev. Earth Planet. Sci.}\
  }\textbf {\bibinfo {volume} {31}},\ \bibinfo {pages} {75} (\bibinfo {year}
  {2003})}\BibitemShut {NoStop}%
\bibitem [{\citenamefont {Montgomery}\ \emph {et~al.}(2006)\citenamefont
  {Montgomery}, \citenamefont {Nicholls}, \citenamefont {Cram},\ and\
  \citenamefont {Saunders}}]{Montgomeryetal2006}%
  \BibitemOpen
  \bibfield  {author} {\bibinfo {author} {\bibfnamefont {M.~T.}\ \bibnamefont
  {Montgomery}}, \bibinfo {author} {\bibfnamefont {M.~E.}\ \bibnamefont
  {Nicholls}}, \bibinfo {author} {\bibfnamefont {T.~A.}\ \bibnamefont {Cram}},\
  and\ \bibinfo {author} {\bibfnamefont {A.~B.}\ \bibnamefont {Saunders}},\
  }\href@noop {} {\bibfield  {journal} {\bibinfo  {journal} {J. Atmos. Sci.}\
  }\textbf {\bibinfo {volume} {63}},\ \bibinfo {pages} {355} (\bibinfo {year}
  {2006})}\BibitemShut {NoStop}%
\bibitem [{\citenamefont {Dunkerton}\ \emph {et~al.}(2009)\citenamefont
  {Dunkerton}, \citenamefont {Montgomery},\ and\ \citenamefont
  {Wang}}]{Dunkertonetal2009}%
  \BibitemOpen
  \bibfield  {author} {\bibinfo {author} {\bibfnamefont {T.~J.}\ \bibnamefont
  {Dunkerton}}, \bibinfo {author} {\bibfnamefont {M.~T.}\ \bibnamefont
  {Montgomery}},\ and\ \bibinfo {author} {\bibfnamefont {Z.}~\bibnamefont
  {Wang}},\ }\href {https://doi.org/10.5194/acp-9-5587-2009} {\bibfield
  {journal} {\bibinfo  {journal} {Atmos. Chem. Phys.}\ }\textbf {\bibinfo
  {volume} {9}},\ \bibinfo {pages} {5587} (\bibinfo {year} {2009})}\BibitemShut
  {NoStop}%
\bibitem [{\citenamefont {Montgomery}\ and\ \citenamefont
  {Smith}(2017)}]{MontgomerySmith2017}%
  \BibitemOpen
  \bibfield  {author} {\bibinfo {author} {\bibfnamefont {M.~T.}\ \bibnamefont
  {Montgomery}}\ and\ \bibinfo {author} {\bibfnamefont {R.~K.}\ \bibnamefont
  {Smith}},\ }\href {https://doi.org/10.1146/annurev-fluid-010816-060022}
  {\bibfield  {journal} {\bibinfo  {journal} {Annu. Rev. Fluid Mech.}\ }\textbf
  {\bibinfo {volume} {49}},\ \bibinfo {pages} {541} (\bibinfo {year}
  {2017})}\BibitemShut {NoStop}%
\bibitem [{\citenamefont {Landau}\ and\ \citenamefont
  {Lifshitz}(1976)}]{LandauLifshitz1976}%
  \BibitemOpen
  \bibfield  {author} {\bibinfo {author} {\bibfnamefont {L.~D.}\ \bibnamefont
  {Landau}}\ and\ \bibinfo {author} {\bibfnamefont {E.~M.}\ \bibnamefont
  {Lifshitz}},\ }\href@noop {} {\emph {\bibinfo {title} {Mechanics}}},\
  \bibinfo {edition} {3rd}\ ed.\ (\bibinfo  {publisher} {Elsevier},\ \bibinfo
  {year} {1976})\BibitemShut {NoStop}%
\bibitem [{\citenamefont {Cambon}\ \emph {et~al.}(1997)\citenamefont {Cambon},
  \citenamefont {Mansour},\ and\ \citenamefont {Godeferd}}]{Cambonetal1997}%
  \BibitemOpen
  \bibfield  {author} {\bibinfo {author} {\bibfnamefont {C.}~\bibnamefont
  {Cambon}}, \bibinfo {author} {\bibfnamefont {N.}~\bibnamefont {Mansour}},\
  and\ \bibinfo {author} {\bibfnamefont {F.}~\bibnamefont {Godeferd}},\
  }\href@noop {} {\bibfield  {journal} {\bibinfo  {journal} {J. Fluid Mech.}\
  }\textbf {\bibinfo {volume} {337}},\ \bibinfo {pages} {303} (\bibinfo {year}
  {1997})}\BibitemShut {NoStop}%
\bibitem [{\citenamefont {Nastrom}\ and\ \citenamefont
  {Gage}(1985)}]{NastromGage1985}%
  \BibitemOpen
  \bibfield  {author} {\bibinfo {author} {\bibfnamefont {G.~D.}\ \bibnamefont
  {Nastrom}}\ and\ \bibinfo {author} {\bibfnamefont {K.~S.}\ \bibnamefont
  {Gage}},\ }\href
  {https://doi.org/10.1175/1520-0469(1985)042<0950:ACOAWS>2.0.CO;2} {\bibfield
  {journal} {\bibinfo  {journal} {J. Atmos. Sci.}\ }\textbf {\bibinfo {volume}
  {42}},\ \bibinfo {pages} {950} (\bibinfo {year} {1985})}\BibitemShut
  {NoStop}%
\bibitem [{\citenamefont {Charney}(1971)}]{Charney1971}%
  \BibitemOpen
  \bibfield  {author} {\bibinfo {author} {\bibfnamefont {J.~G.}\ \bibnamefont
  {Charney}},\ }\href@noop {} {\bibfield  {journal} {\bibinfo  {journal} {J.
  Atmos. Sci.}\ }\textbf {\bibinfo {volume} {28}},\ \bibinfo {pages} {1087}
  (\bibinfo {year} {1971})}\BibitemShut {NoStop}%
\bibitem [{\citenamefont {Ding}\ \emph {et~al.}(2026)\citenamefont {Ding},
  \citenamefont {Bobas}, \citenamefont {Scolan}, \citenamefont {Young},\ and\
  \citenamefont {Read}}]{Dingetal2026}%
  \BibitemOpen
  \bibfield  {author} {\bibinfo {author} {\bibfnamefont {S.-S.}\ \bibnamefont
  {Ding}}, \bibinfo {author} {\bibfnamefont {H.}~\bibnamefont {Bobas}},
  \bibinfo {author} {\bibfnamefont {H.}~\bibnamefont {Scolan}}, \bibinfo
  {author} {\bibfnamefont {R.~M.~B.}\ \bibnamefont {Young}},\ and\ \bibinfo
  {author} {\bibfnamefont {P.~L.}\ \bibnamefont {Read}},\ }\href
  {https://doi.org/10.1103/n2nj-dg5r} {\bibfield  {journal} {\bibinfo
  {journal} {Phys. Rev. Lett.}\ }\textbf {\bibinfo {volume} {136}},\ \bibinfo
  {pages} {114101} (\bibinfo {year} {2026})}\BibitemShut {NoStop}%
\bibitem [{\citenamefont {Lindborg}\ and\ \citenamefont
  {Cho}(2001)}]{LindborgCho2001}%
  \BibitemOpen
  \bibfield  {author} {\bibinfo {author} {\bibfnamefont {E.}~\bibnamefont
  {Lindborg}}\ and\ \bibinfo {author} {\bibfnamefont {J.~Y.~N.}\ \bibnamefont
  {Cho}},\ }\href@noop {} {\bibfield  {journal} {\bibinfo  {journal} {J.
  Geophys. Res.}\ }\textbf {\bibinfo {volume} {106}},\ \bibinfo {pages} {10233}
  (\bibinfo {year} {2001})}\BibitemShut {NoStop}%
\bibitem [{\citenamefont {Galperin}\ \emph {et~al.}(2025)\citenamefont
  {Galperin}, \citenamefont {Nickerson}, \citenamefont {King},\ and\
  \citenamefont {Zhang}}]{Galperinetal2025}%
  \BibitemOpen
  \bibfield  {author} {\bibinfo {author} {\bibfnamefont {B.}~\bibnamefont
  {Galperin}}, \bibinfo {author} {\bibfnamefont {A.~K.}\ \bibnamefont
  {Nickerson}}, \bibinfo {author} {\bibfnamefont {G.~P.}\ \bibnamefont
  {King}},\ and\ \bibinfo {author} {\bibfnamefont {J.~A.}\ \bibnamefont
  {Zhang}},\ }\href {https://doi.org/10.1029/2024JD042733}
  {\bibfield  {journal} {\bibinfo  {journal} {J. Geophys. Res.}\ }\textbf
  {\bibinfo {volume} {130}},\ \bibinfo {pages} {e2024JD042733} (\bibinfo {year}
  {2025})}\BibitemShut {NoStop}%
\bibitem [{\citenamefont {Sukoriansky}\ and\ \citenamefont
  {Galperin}(2016)}]{SukorianskyGalperin2016}%
  \BibitemOpen
  \bibfield  {author} {\bibinfo {author} {\bibfnamefont {S.}~\bibnamefont
  {Sukoriansky}}\ and\ \bibinfo {author} {\bibfnamefont {B.}~\bibnamefont
  {Galperin}},\ }\href {https://doi.org/10.1017/jfm.2016.568} {\bibfield
  {journal} {\bibinfo  {journal} {J. Fluid Mech.}\ }\textbf {\bibinfo {volume}
  {805}},\ \bibinfo {pages} {384} (\bibinfo {year} {2016})}\BibitemShut
  {NoStop}%
\bibitem [{\citenamefont {Galperin}\ \emph {et~al.}(2021)\citenamefont
  {Galperin}, \citenamefont {Sukoriansky},\ and\ \citenamefont
  {Qiu}}]{Galperinetal2021}%
  \BibitemOpen
  \bibfield  {author} {\bibinfo {author} {\bibfnamefont {B.}~\bibnamefont
  {Galperin}}, \bibinfo {author} {\bibfnamefont {S.}~\bibnamefont
  {Sukoriansky}},\ and\ \bibinfo {author} {\bibfnamefont {B.}~\bibnamefont
  {Qiu}},\ }\href {https://doi.org/10.1007/s10236-021-01444-1} {\bibfield
  {journal} {\bibinfo  {journal} {Ocean Dyn.}\ }\textbf {\bibinfo {volume}
  {71}},\ \bibinfo {pages} {475} (\bibinfo {year} {2021})}\BibitemShut
  {NoStop}%
\bibitem [{\citenamefont {Galperin}\ and\ \citenamefont
  {Sukoriansky}(2020)}]{GalperinSukoriansky2020}%
  \BibitemOpen
  \bibfield  {author} {\bibinfo {author} {\bibfnamefont {B.}~\bibnamefont
  {Galperin}}\ and\ \bibinfo {author} {\bibfnamefont {S.}~\bibnamefont
  {Sukoriansky}},\ }\href {https://doi.org/10.1103/PhysRevFluids.5.063803}
  {\bibfield  {journal} {\bibinfo  {journal} {Phys. Rev. Fluids}\ }\textbf
  {\bibinfo {volume} {5}},\ \bibinfo {eid} {063803} (\bibinfo {year}
  {2020})}\BibitemShut {NoStop}%
\bibitem [{\citenamefont {Pielke}\ and\ \citenamefont
  {Landsea}(1998)}]{PielkeLandsea1998}%
  \BibitemOpen
  \bibfield  {author} {\bibinfo {author} {\bibfnamefont {R.~A.}\ \bibnamefont
  {Pielke}}\ and\ \bibinfo {author} {\bibfnamefont {C.~W.}\ \bibnamefont
  {Landsea}},\ }\href@noop {} {\bibfield  {journal} {\bibinfo  {journal}
  {Weather Forecast.}\ }\textbf {\bibinfo {volume} {13}},\ \bibinfo {pages}
  {621} (\bibinfo {year} {1998})}\BibitemShut {NoStop}%
\bibitem [{\citenamefont {Emanuel}(1986)}]{Emanuel1986}%
  \BibitemOpen
  \bibfield  {author} {\bibinfo {author} {\bibfnamefont {K.~A.}\ \bibnamefont
  {Emanuel}},\ }\href
  {https://doi.org/10.1175/1520-0469(1986)043<0585:AASITF>2.0.CO;2} {\bibfield
  {journal} {\bibinfo  {journal} {J. Atmos. Sci.}\ }\textbf {\bibinfo {volume}
  {43}},\ \bibinfo {pages} {585} (\bibinfo {year} {1986})}\BibitemShut
  {NoStop}%
\bibitem [{\citenamefont {Hersbach}\ \emph {et~al.}(2020)\citenamefont
  {Hersbach}, \citenamefont {Bell}, \citenamefont {Berrisford} \emph
  {et~al.}}]{Hersbachetal2020}%
  \BibitemOpen
  \bibfield  {author} {\bibinfo {author} {\bibfnamefont {H.}~\bibnamefont
  {Hersbach}}, \bibinfo {author} {\bibfnamefont {B.}~\bibnamefont {Bell}},
  \bibinfo {author} {\bibfnamefont {P.}~\bibnamefont {Berrisford}}, \emph
  {et~al.},\ }\href {https://doi.org/10.1002/qj.3803} {\bibfield  {journal}
  {\bibinfo  {journal} {Q. J. R. Meteorol. Soc.}\ }\textbf {\bibinfo {volume}
  {146}},\ \bibinfo {pages} {1999} (\bibinfo {year} {2020})}\BibitemShut
  {NoStop}%
\bibitem [{\citenamefont {B{\"u}hler}\ \emph {et~al.}(2014)\citenamefont
  {B{\"u}hler}, \citenamefont {Callies},\ and\ \citenamefont
  {Ferrari}}]{Buhleretal2014}%
  \BibitemOpen
  \bibfield  {author} {\bibinfo {author} {\bibfnamefont {O.}~\bibnamefont
  {B{\"u}hler}}, \bibinfo {author} {\bibfnamefont {J.}~\bibnamefont
  {Callies}},\ and\ \bibinfo {author} {\bibfnamefont {R.}~\bibnamefont
  {Ferrari}},\ }\href@noop {} {\bibfield  {journal} {\bibinfo  {journal} {J.
  Fluid Mech.}\ }\textbf {\bibinfo {volume} {756}},\ \bibinfo {pages} {1007}
  (\bibinfo {year} {2014})}\BibitemShut {NoStop}%
\bibitem [{\citenamefont {Craya}(1958)}]{Craya1958}%
  \BibitemOpen
  \bibfield  {author} {\bibinfo {author} {\bibfnamefont {A.}~\bibnamefont
  {Craya}},\ }\href@noop {} {\emph {\bibinfo {title} {Contribution {\`a}
  l'analyse de la turbulence associ{\'e}e {\`a} des vitesses moyennes}}},\
  \bibinfo {type} {Publications Scientifiques et Techniques}\ \bibinfo {number}
  {345}\ (\bibinfo  {institution} {Minist{\`e}re de l'Air},\ \bibinfo {address}
  {Paris},\ \bibinfo {year} {1958})\BibitemShut {NoStop}%
\bibitem [{\citenamefont {Herring}(1974)}]{Herring1974}%
  \BibitemOpen
  \bibfield  {author} {\bibinfo {author} {\bibfnamefont {J.~R.}\ \bibnamefont
  {Herring}},\ }\href@noop {} {\bibfield  {journal} {\bibinfo  {journal} {Phys.
  Fluids}\ }\textbf {\bibinfo {volume} {17}},\ \bibinfo {pages} {859} (\bibinfo
  {year} {1974})}\BibitemShut {NoStop}%
\bibitem [{\citenamefont {Monin}\ and\ \citenamefont
  {Yaglom}(1975)}]{MoninYaglom1975}%
  \BibitemOpen
  \bibfield  {author} {\bibinfo {author} {\bibfnamefont {A.~S.}\ \bibnamefont
  {Monin}}\ and\ \bibinfo {author} {\bibfnamefont {A.~M.}\ \bibnamefont
  {Yaglom}},\ }\href@noop {} {\emph {\bibinfo {title} {Statistical Fluid
  Mechanics}}}\ (\bibinfo  {publisher} {MIT Press},\ \bibinfo {year}
  {1975})\BibitemShut {NoStop}%
\bibitem [{SM()}]{SM}%
  \BibitemOpen
  \href@noop {} {}\bibinfo {note} {See Supplemental Material at [URL will be
  inserted by publisher] for the isotropy-free analysis and the tests of scale
  range, sampling uncertainty and storm masking.}\BibitemShut {Stop}%
\bibitem [{\citenamefont {Xu}\ and\ \citenamefont {Fu}(2012)}]{XuFu2012}%
  \BibitemOpen
  \bibfield  {author} {\bibinfo {author} {\bibfnamefont {Y.}~\bibnamefont
  {Xu}}\ and\ \bibinfo {author} {\bibfnamefont {L.-L.}\ \bibnamefont {Fu}},\
  }\href {https://doi.org/10.1175/JPO-D-12-0106.1} {\bibfield  {journal}
  {\bibinfo  {journal} {J. Phys. Oceanogr.}\ }\textbf {\bibinfo {volume}
  {42}},\ \bibinfo {pages} {2229} (\bibinfo {year} {2012})}\BibitemShut
  {NoStop}%
\bibitem [{\citenamefont {Chavas}\ and\ \citenamefont
  {Reed}(2019)}]{ChavasReed2019}%
  \BibitemOpen
  \bibfield  {author} {\bibinfo {author} {\bibfnamefont {D.~R.}\ \bibnamefont
  {Chavas}}\ and\ \bibinfo {author} {\bibfnamefont {K.~A.}\ \bibnamefont
  {Reed}},\ }\href {https://doi.org/10.1175/JAS-D-19-0001.1} {\bibfield
  {journal} {\bibinfo  {journal} {J. Atmos. Sci.}\ }\textbf {\bibinfo {volume}
  {76}},\ \bibinfo {pages} {2257} (\bibinfo {year} {2019})}\BibitemShut
  {NoStop}%
\bibitem [{\citenamefont {Emanuel}\ and\ \citenamefont
  {Nolan}(2004)}]{EmanuelNolan2004}%
  \BibitemOpen
  \bibfield  {author} {\bibinfo {author} {\bibfnamefont {K.~A.}\ \bibnamefont
  {Emanuel}}\ and\ \bibinfo {author} {\bibfnamefont {D.~S.}\ \bibnamefont
  {Nolan}},\ }in\ \href@noop {} {\emph {\bibinfo {booktitle} {Preprints, 26th
  Conf. on Hurricanes and Tropical Meteorology}}}\ (\bibinfo  {publisher}
  {Amer. Meteor. Soc.},\ \bibinfo {address} {Miami, FL},\ \bibinfo {year}
  {2004})\ pp.\ \bibinfo {pages} {240--241}\BibitemShut {NoStop}%
\end{thebibliography}%
